\documentclass[journal,twocolumn]{IEEEtran}
\usepackage{cite}
\usepackage{amsmath,amsfonts,amsxtra,amssymb,latexsym,amscd,amsthm,mathrsfs,bm}
\usepackage{mathtools}
\usepackage{tabularx}
\usepackage{graphicx}
\usepackage{hyperref}
\usepackage{xcolor}
\usepackage{algorithm}
\usepackage{algorithmic}
\usepackage{cite}
\usepackage[colorinlistoftodos,textsize=tiny]{todonotes}
\usepackage{balance}
\usepackage{epstopdf}
\usepackage{balance}
\usepackage{booktabs}
\usepackage{lipsum}

\usepackage{caption}
\usepackage{subcaption}

\newcommand{\nn}{\nonumber}

\usepackage{eqparbox}

\usepackage{etoolbox}  % patch def of algorithmic environment
\makeatletter
\patchcmd{\algorithmic}{\addtolength{\ALC@tlm}{\leftmargin} }{\addtolength{\ALC@tlm}{\leftmargin}}{}{}
\makeatother

\newcommand{\T}{\mathsf{T}}

\newcommand{\Hi}{\mathsf{H}}

\title{\Large PAPR Reduction for AFDM  by Affine-Domain Circular Shift without Side Information
	\thanks{T. K. Nguyen and E. Bedeer are with the Department of Electrical and Computer Engineering, University of Saskatchewan, Saskatoon, Canada S7N5A9. Emails: \{khai.nguyen,  e.bedeer\}@usask.ca.}
	\thanks{	This work was supported by the NSERC/Cisco Industrial Research
		Chair.}
}

\author{\IEEEauthorblockN{The Khai Nguyen and Ebrahim Bedeer}
	\vspace{-0.8cm}}

\begin{document}
	
	\maketitle
	
	\begin{abstract}
		This paper proposes a novel method to reduce the peak-to-average-power-ratio (PAPR) of affine frequency division multiplexing (AFDM) signals without side information (SI). The method is based on circularly shifting the original transmit signal in the affine domain, and selecting the shifted candidate with the lowest PAPR. Next, a maximum-likelihood-based (MLB)  receiver is derived, which exploits the position of the AFDM pilot and guard band to detect the shift applied at the transmitter  without SI.  Simulation results show that the proposed method can achieve 2.5 to 4 dB in PAPR reduction as compared to original AFDM and existing schemes, which can be translated into significantly lower error rate, depending on the quality of power amplifiers.
	\end{abstract}
	\vspace*{-0.2cm}
	
	\section{Introduction}\label{Sec:Intro}
	Affine frequency division multiplexing (AFDM) is a novel multicarrier  technique that has gained increasing attentions recently \cite{Bemani2023,Zheng2025}. AFDM uses chirp-based waveforms derived from the discrete affine Fourier transform (DAFT), enabling joint time-frequency spreading of data symbols. Unlike orthogonal frequency division multiplexing (OFDM), which is susceptible to inter-carrier-interference due to Doppler effect and has poor  diversity in doubly selective channel, AFDM can maintain full delay-Doppler diversity even under severe Doppler shift, making it ideal for high-mobility scenarios such as vehicular or satellite communications \cite{Bemani2023,Zheng2025}. 
	However, like OFDM, AFDM  exhibits high peak-to-average power ratio (PAPR),  which motivates ongoing research on novel PAPR reduction methods suitable to AFDM.
	
	In \cite{Yuan2025}, the authors  propose to vary the pre-chirp parameter $c_2$ across subcarrier groups in a non-enumerated way, and select the candidate with the lowest peak power. This method can achieve up to 5 dB lower PAPR  than original AFDM, at the cost of higher  complexity and rate loss due to side information (SI). In \cite{Reddy2025} the authors utilize $\mu$-law companding to compress signal amplitude  and limit peak power, while maintaining the Doppler resilience of AFDM. The paper derives an upper bound for PAPR and validates the approach through simulation, showcasing   significant PAPR reduction  compared to original AFDM (up to 4 dB). The work in \cite{Ali2026} proposes a unified pre-modulation data-spreading algorithm to reduce the PAPR in AFDM  systems without SI. This method uses different well-known transformations to redistribute signal energy before modulation, which demonstrates up to 4 dB PAPR reduction while maintaining spectral efficiency. 
	In \cite{Reddy2026} a hybrid approach is introduced, which  integrates  Zadoff-Chu sequence matrix (ZCM) precoding with normalized $\mu$-law and $A$-law companding to compress signal peaks while maintaining  delay-Doppler diversity. The reported results  show a moderate PAPR reduction and over 0.5 dB signal-to-noise ratio (SNR) improvement over conventional AFDM. 
	
	Despite the promising performances, the aforementioned works  overlook the pilot structure of AFDM and how channel estimation could be affected \cite{Huang2026}. To address this,  the authors in \cite{Huang2026} propose to replace the AFDM guard intervals  with low-magnitude samples and jointly perform active constellation extension on data symbols, which is trained by a neural network using PAPR as the loss function. Although the non-zero guard band worsens channel estimation quality, the reported results show a notable PAPR reduction with minimal impact on bit error rate (BER),  at the expense of  higher complexity.
	The work in \cite{Choi2026}  selects the optimal value of $c_2$ that yields the lowest PAPR, from a predetermined set. The candidate set is chosen such that the applied $c_2$ value can be recognized at the receiver without SI, and does not affect the pilot structure of AFDM. The proposed approach was shown to achieve  2-3 dB PAPR reduction  while  improving BER performance compared to existing PAPR reduction schemes for AFDM.
	
	Against the aforementioned background, in this paper, we propose a  PAPR reduction method utilizing  circular shift in the affine domain.   The proposed method searches for the optimal circular shift that results in the lowest peak power among the candidate set. After that, a maximum-likelihood-based (MLB) blind detector is derived, which exploits the pilot structure of AFDM  to detect the number of samples by which the original signal  was circularly shifted  at the transmitter, without SI. 
	Next,   we propose to limit the number of shifted candidates by applying a minimum spacing between any two different shifts, which helps  reduce the error of the  MLB blind detector due to the similarity between close  candidates.
	%Furthermore, we also show that the calculation of the time domain AFDM signals associated with different candidates can be done with a circular convolution of the original time-domain signal with a sparse  vector with mostly low amplitude samples, which can significantly reduce the complexity order of the proposed method.
	
	The remainder of this paper is organized as follows. Section~\ref{Sec:sys} presents the  system model and the proposed PAPR reduction method with affine-domain circular shift. Section~\ref{Sec:blind} derives the blind estimation for the shift without SI. Section~\ref{Sec:com} presents complexity analysis. Then Section~\ref{Sec:sim}  provides simulation results. Finally, Section~\ref{Sec:con}  concludes the paper.
	
	\textit{Notations:} The pairwise multiplication is  $\otimes$. The circular convolution is  $\odot$. Transposed and Hermitian matrices are denoted by the superscript $\T$ and $\Hi$, respectively. Identity matrix of size $N$ is $\mathbf{I}_N$. The modulo operator is  $\mathrm{mod}$. The set of $d_1\times d_2$ complex matrices is  $\mathbb{C}^{d_1\times d_2}$. The central complex normal distribution with variance $\sigma^2$ is denoted by $\mathcal{CN}\left(0,\sigma^2\right)$. 	The circular shifts to the left and right are denoted by $<<$ and $>>$, respectively. The determinant of a matrix is  $\mathrm{det}(\cdot)$.

	\section{PAPR Reduction by Affine-Domain Circular Shift}\label{Sec:sys}
	\subsection{System model}
	Consider an AFDM system with $N$ subcarriers and the parameters $c_1$, $c_2$. The input signal to the system in the affine domain $\boldsymbol{x}=[x[0], x[1]\dots x[N-1]]^{\T}\in\mathbb{C}^{N\times 1}$ is drawn from an $M$-QAM constellation. This signal is AFDM modulated into the time-domain signal  $\boldsymbol{s}=[s[0],s[1]\dots s[N-1]]^{\T}$ as:
	\begin{equation}
		s[n] = \sum_{m=0}^{N-1}\frac{x[m]}{\sqrt{N}}\exp\left\{j2\pi\left[c_1n^2 + \frac{mn}{N} + c_2m^2\right]\right\}.
	\end{equation}
	This can be rewritten in the matrix form as:
	\begin{equation}\label{eq.AFDM}
		\boldsymbol{s}= \mathbf{A}^{\Hi}\boldsymbol{x} = \mathbf{\Lambda}_1^{\Hi}\mathbf{F}^{-1}\mathbf{\Lambda}_2^{\Hi}\boldsymbol{x},
	\end{equation}	
	where $\mathbf{\Lambda}_i$ = $\mathrm{diag}(\exp(-j2\pi c_i n^2))$, (for $i\in\{1,2\}$),	and $\mathbf{F}$ is the unitary discrete Fourier transform (DFT) matrix.
	Then, a chirp periodic prefix (CPP) is added to avoid inter-symbol-interference (ISI). After that, the  signal is transmitted over a doubly selective channel with impulse response:
	\begin{equation}
		h[n,\tau] = \sum_{\ell=0}^{L-1} h_{\ell}\delta(\tau - \tau_{\ell})\exp\left\{j2\pi f_{\ell} n\right\},
	\end{equation}
		where $L$ is the number of paths, $h_{\ell}\sim\mathcal{CN}(0,\rho_{\ell})$ is the complex channel gain with variance $\rho_{\ell}$, while $\tau_{\ell}$, $f_{\ell}$ are the  delay  and  Doppler shift   of the ${\ell}$th path  normalized by the sampling period. Without the loss of generality, we set $\tau_{\ell}=\ell-1$, which means that the maximum delay is $\max(\tau_{\ell})=L-1$. With the use of a 	CPP to avoid ISI, the input-output signal relationship of AFDM signals over the channel 	 after CPP removal is \cite{Bemani2023}:
	\begin{equation}
		\boldsymbol{r} = \mathbf{H}\boldsymbol{s} + \boldsymbol{w},
	\end{equation}
	where $\boldsymbol{w}\sim\mathcal{CN}(0,\sigma^2\mathbf{I}_{N})$ is AWGN noise and the channel matrix $\mathbf{H}$ is defined as
	$
	\mathbf{H}\triangleq\sum_{{\ell}=1}^{L}h_{\ell}\mathbf{\Gamma}_p\mathbf{\Delta}_{f_{\ell}}\mathbf{\Pi}^{\tau_{\ell}},
	$
	where 
	\begin{equation}
		\small
		\mathbf{\Pi} = \begin{bmatrix}
			&0\;&0\;&\dots\;&0\; &1\;&\\
			&1\;&0\;&\dots\;&0\;&0\;&\\
			&\vdots\;&\ddots\;&\ddots&\ddots &\vdots\;&\\
			&0&0&\dots&1&0\;&
		\end{bmatrix},
	\end{equation} $\boldsymbol{\Delta}_{f_\ell}=\mathrm{diag}\{\exp\left\{j2\pi f_{\ell} n\right\}\}$, ($n=0,1\dots N-1$) and $\mathbf{\Gamma}_p$ is an identity matrix when $c_1$ is set as \cite{Bemani2023}:
	\begin{equation}\label{eq.c1}
		c_1 =\frac{2(\alpha_{\mathrm{max}} +\xi)+1}{N},
	\end{equation} 
	where the non-negative integer	 $\xi$ is the  spacing parameter.
	The received signal is converted back to the affine domain by \cite{Bemani2023}:
	\begin{equation}
		\boldsymbol{y} = \mathbf{A}\boldsymbol{r}= \mathbf{H}_{\mathrm{eff}}\boldsymbol{x} + \tilde{\boldsymbol{w}},
	\end{equation}
	where 
	$
	\mathbf{H}_{\mathrm{eff}} \triangleq \sum_{{\ell}=0}^{L-1}h_{\ell}\mathbf{H}_{\ell}
	$ and $\mathbf{H}_{\ell}\triangleq\mathbf{A}\mathbf{\Gamma}_{\ell}\mathbf{\Delta}_{f_{\ell}}\mathbf{\Pi}^{\tau_{\ell}}\mathbf{A}^{\Hi}$.
	Let the Doppler normalized to subcarrier spacing be defined as $\nu_{\ell}\triangleq Nf_{\ell}=\alpha_{\ell} + a_{\ell}$, where $\alpha_{\ell}$ is the integer part $\alpha_{\ell}\in[-\alpha_{\mathrm{max}},\alpha_{\mathrm{max}}]$ and $a_{\ell}$ is the fractional part that satisfies $-0.5<a_{\ell}\leq0.5$. With $c_1$ 	defined as in \eqref{eq.c1},
	the full diversity of the channel is achieved \cite{Bemani2023}. The  spacing parameter $\xi$ is responsible for separating the channels  of different paths in $\mathbf{H}_{\mathrm{eff}}$ to avoid overlapping among different paths (different $\mathbf{H}_{\ell}$) that can degrade the diversity (with integer Doppler, $\xi=0$).
	\subsection{PAPR reduction with affine domain circular shift}
	
	The PAPR of the AFDM modulated signal is defined as:
	\begin{equation}
		\mathrm{PAPR}\left\{\boldsymbol{s}\right\} = \frac{\mathrm{max}(|s[n]|^2)}{\mathbb{E}\left\{|s[n]|^2\right\}}.
	\end{equation}
		Let $	\tilde{\boldsymbol{s}}_0 =  \mathbf{F}^{-1}\mathbf{\Lambda}_2^{\Hi}\boldsymbol{x}=\mathrm{IDFT}\left\{\boldsymbol{a}_2\otimes\boldsymbol{x}\right\}$, where $\boldsymbol{a}_2=\mathrm{diag}(\boldsymbol{\Lambda}_2^{\Hi}) = \exp(j2\pi c_2 n^2)$.
		From \eqref{eq.AFDM}, it can be seen that $\boldsymbol{s}=\mathbf{\Lambda}_1^{\Hi}\tilde{\boldsymbol{s}}_0$, and thus, the multiplication with $\mathbf{\Lambda}_1^{\Hi}$ simply makes $\boldsymbol{s}$  a phase rotation of $\tilde{\boldsymbol{s}}_0$ , which does not affect the sample power. Thus, it can be seen that:
		\begin{equation}
			\mathrm{PAPR}\left\{{\boldsymbol{s}}\right\}=	\mathrm{PAPR}\left\{\tilde{\boldsymbol{s}}_0\right\}.
	\end{equation}
	This simplifies the PAPR reduction problem since the parameter $c_1$ is removed.
	Then, we propose to transmit a circularly shifted version of $\boldsymbol{x}$ that has a lower peak power, instead of the original 	$\boldsymbol{x}$. The search for the optimal shift  will resort to choosing the the one leading to the lower peak power:
	\begin{align}\label{eq.or_op}
		\hat{k} &= \underset{k=0,1\dots N-1}{\mathrm{argmin}}\quad \mathrm{max}\left\{|\mathrm{IDFT}\left\{\boldsymbol{a}_2\otimes(\boldsymbol{x}>> k)\right\}|\right\} 
		%&=\underset{k=0,1\dots N-1}{\mathrm{argmin}}\quad \mathrm{max}\left\{\mathrm{IFFT}\left\{(\boldsymbol{a}_2<<k)\otimes\boldsymbol{x}\right\}\right\}.
	\end{align}
	%With the exhaustive search approach, each shift version will require the recalculation of the IFFT. Consequentially, the complexity of this approach is $\mathcal{O}(N^2\mathrm{log}_2 N)$.
	Then, the signal ${\tilde{\boldsymbol{x}}_{\hat{k}}}\triangleq(\boldsymbol{x}>>\hat{k})$ will be AFDM modulated by \eqref{eq.AFDM} and transmitted over the channel instead of $\boldsymbol{x}$. 
	
	In  OFDM, a circular shift  in the frequency domain only  results in  a phase shift  in the time-domain signal, which does not change the PAPR. By contrast, in AFDM,   circularly shifting $\boldsymbol{x}$ before prechirping can change  the PAPR of the AFDM in the time domain due to the multiplication with $\mathbf{\Lambda}_2^{\Hi}$ in \eqref{eq.AFDM}. 
	With this method, the receiver needs to know the  shift applied at the transmitter to revert the signal in the affine domain back to original. The use of SI as in many existing works in PAPR reduction is not applicable, since SI can only be extracted after the data has been detected, which is not possible without knowing how much $\boldsymbol{x}$ has been shifted first. The channel estimation is also infeasible for the same reason. Thus, in the next section, we propose a blind estimation for the shift in the affine domain, without SI.

	\section{Blind Estimation of Affine Domain Shift}\label{Sec:blind}
	%	This section proposes a method to blindly detect the shift $\hat{k}$ that has been performed at the transmitter on $\boldsymbol{x}$, without requiring any SI overhead. The blind estimation is important for two reasons. First, it is obvious that if the receiver does not know how many samples $\boldsymbol{x}$ has been circularly shifted, even if all the QAM symbols carried on the chirps has been detected, they cannot be placed in the correct original order. Second, one important requirement in AFDM  is the channel estimation.   If the receiver does not know the position of the pilot (due to circular shift), channel estimation cannot be performed. 
	This section proposes a blind estimation method for the shift associated with PAPR reduction. Conventionally, AFDM requires a single pilot sample embedded in $\boldsymbol{x}$, surrounded by $Q$ zero guard samples on each side of the pilot  \cite{Bemani2023}, where
	\begin{equation}
		Q=(\max(\tau_{\ell})+1)\times\beta - 1=L\beta -1,
	\end{equation} 
	with 
	$
	\beta \triangleq 2\alpha_{\xi} +1
	$, and $\alpha_{\xi}=\alpha_{\mathrm{max}}+\xi$.
	The pilot embedded  signal in the affine domain is therefore given as:
	\begin{equation}\label{eq.pilot}
		\boldsymbol{x}=\left[P,0\dots0,x[Q+1]\dots x[N-Q-1],0\dots0\right],
	\end{equation}
	where $P$ is the pilot symbol. 
	
\begin{figure}[t!]
		\centering
		\vspace{-.45cm}
		\includegraphics[width=0.4\textwidth]{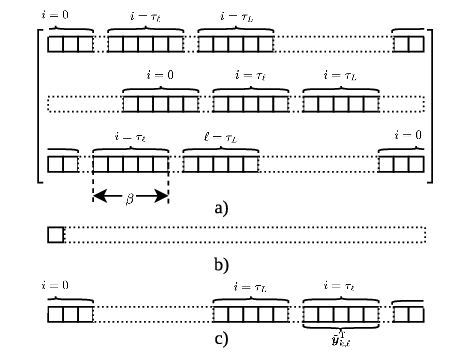}
		\caption{Structure of: (a) $\mathbf{H}_{\mathrm{eff}}$, (b) $\boldsymbol{x}^{\T}$, and (c) $(\mathbf{H}_{\mathrm{eff}}\boldsymbol{x})^{\T}$.}
		\label{fig-bin-pilot2}
	\end{figure}

	Next, the effect of the channel $\mathbf{H}_{\mathrm{eff}}$ on the pilot-embedded signal $\boldsymbol{x}$ needs to be examined. According to \cite{Bemani2023}, by setting $\xi$ properly, $\mathbf{H}_{\mathrm{eff}}$ has a special structure where each row (and column) is composed of $\max(\tau_{\ell})+1=L$ adjacent segments, each corresponds to a different delay (from 0 to $L-1$) and has the length of $\beta$ samples as shown in Fig.~\ref{fig-bin-pilot2}-(a).
	For the simplicity in explaining the effect of $\mathbf{H}_{\mathrm{eff}}$ on $\boldsymbol{x}$, assume that no data is transmitted ($
	\boldsymbol{x}=\left[P,0,0\dots0\right]
	$ as in Fig.~\ref{fig-bin-pilot2}-(b)) and the receiver is noise-free.
	In this case, the received signal $\boldsymbol{y}=\mathbf{H}_{\mathrm{eff}}\boldsymbol{x}$ will be exactly the first column of $\mathbf{H}_{\mathrm{eff}}$, which also has the special structure with $L$ adjacent segments of $\beta$ samples, while the remaining samples are zeros as shown in Fig.~\ref{fig-bin-pilot2}-(c). %In specific, according to Fig.~\ref{fig-bin-pilot2}, the segment associated with the  $\ell$th path contains the sample from $y$. 
	These $L$ segments total $L\times\beta$ samples, and is confined completely within the guard interval of $2Q+1$ samples of $\boldsymbol{x}$. More importantly, the remaining  samples of the guard interval, which  does not belong to any  segments of any paths, guarantee that interference from data samples cannot reach any  segments. This means that in a general case with pilot embedded plus data as in \eqref{eq.pilot}, the properties of these $L$ segments analyzed earlier still hold, and are not affected by interference from data \cite{Bemani2023}. This is the most important establishment that enables the blind estimation of the shift for the proposed PAPR reduction.

For the ease in explanation, we will first start with the blind estimation of the shift for integer Doppler case. After that, the non-integer Doppler case can be addressed.
	\subsection{Integer Doppler case}
	As discussed earlier, with the parameters setting in \eqref{eq.c1}, the channel with different delays or Doppler shifts are separated in the affine domain \cite{Bemani2023}. Consequently, when the pilot-embedded signal is transmitted, the $L$ pilot-associated segments of $\boldsymbol{y}=\mathbf{H}_{\mathrm{eff}}\boldsymbol{x}$,  as shown in Fig.~\ref{fig-bin-pilot2}-(c), are mutually independent. Based on this, we will derive a MLB blind detector to determine if a segment of $\beta$ samples taken from the received vector $\boldsymbol{y}$ is associated with a channel path. Note that in the integer Doppler case, $\xi$ is set to zero \cite{Bemani2023}, which leads to $\beta = 2\alpha_{\mathrm{max}}+1$.

	Suppose that  the signal $\tilde{\boldsymbol{x}}_k = {\boldsymbol{x}}>>k$ was transmitted, the position of the pilot will also be shifted by $k$ samples (i.e. $\tilde{x}[k]=P$). Thus, to detect if the original signal $\boldsymbol{x}$ was shifted to the right by $k$ samples, the received signal $\boldsymbol{y}$ will be circularly shifted to the left by $k$ samples to obtain $
	\tilde{\boldsymbol{y}}_k=\boldsymbol{y}<<k
	$.
	%	to revert the pilot to the original position at index $m=0$. Then, the likelihood function that 
	% the $\max(\tau_{\ell})(2\alpha_{\mathrm{max}}+1)$ samples from $\tilde{y}_i[N-Q+\alpha_{\mathrm{max}}]$ to $\tilde{y}_i[N-1]$ and $\tilde{y}_i[0]$ to $\tilde{y}_i[\alpha_{\mathrm{max}}]$ of $\tilde{\boldsymbol{y}}_i$, which are associated with the $L$ pilot-associated segments, will be calculated given $\boldsymbol{x}$ was shifted by $i$ samples to reduce the PAPR.
	%	
	%	
	Let $\tilde{\boldsymbol{y}}_{k,\ell}\in\mathbb{C}^{\beta\times 1}$ be defined as $
	\tilde{\boldsymbol{y}}_{k,\ell}\triangleq [\tilde{y}_k[(-\ell\beta-\alpha_{\xi})\mathrm{mod}N]\dots \tilde{y}_k[(-\ell\beta+\alpha_{\xi})\mathrm{mod}N]]^{\T}
	$,
	which are the $\beta$ samples of the $\ell$-th segment, given that $\tilde{\boldsymbol{x}}_k = \boldsymbol{x}>>k$ was transmitted, as shown in Fig.~\ref{fig-bin-pilot2}-(c). Obviously, when the Doppler associated with this segment  is $\alpha_\ell=0$, the channel of the $\ell$th tap will be at the central position of $\tilde{\boldsymbol{y}}_{k,\ell}$, i.e. $\tilde{y}_k[(-\ell\beta)\mathrm{mod}N]=h_{\ell}$, while other samples are zeros. Whereas, when the Doppler is $\alpha_\ell=\pm1$, the channel response of the $\ell$th path will be shifted to the left and right by one sample, respectively.
	Generally, with integer Doppler, within $\tilde{\boldsymbol{y}}_{k,\ell}$, only one sample can be associated with the channel of the $\ell$th tap $h_{\ell}\sim\mathcal{CN}(0,\rho_{\ell})$, while the  rest are just noise. Mathematically, with the integer Doppler $\alpha_\ell$:
	\begin{align}\label{eq.dis_nonint}
		%\tilde{y}_i[(-p\tilde{\alpha}+\alpha_{p})\mathrm{mod}N]&\sim \mathcal{CN}(0,\rho_p+\sigma^2)\nn\\
		\tilde{\boldsymbol{y}}_{k,\ell}&\sim \mathcal{CN}(0,\mathrm{diag}\{\boldsymbol{g}_{\alpha_{\ell}}\}+\sigma^2\mathbf{I}_{\beta}),\;
	\end{align}
	where $\boldsymbol{g}_{\alpha_{\ell}}=[{g}_{\alpha_{\ell,1}},{g}_{\alpha_{\ell,2}}\dots {g}_{\alpha_{\ell},\beta}]^{\T}$ with all zero elements except  ${g}_{\alpha_{\ell,i}}=\rho_{\ell}$ for $i=\alpha_{\mathrm{max}}+1+\alpha_\ell$. 
	%	\begin{equation}
		%		{g}_{\alpha_{\ell,i}}=\begin{cases}
			%			\rho_{\ell},\;i=\alpha_{\mathrm{max}}+1+\alpha_\ell,\\
			%			0,\;\mathrm{otherwise}.
			%		\end{cases}
		%	\end{equation}
	Consequently,
	the the probability density function (pdf) of $\tilde{\boldsymbol{y}}_k$ conditioned on the integer Doppler shift $\alpha_{\ell}$ is:
	\begin{align}
		&\mathrm{Pr}(\tilde{\boldsymbol{y}}_k|\alpha_{\ell})=\frac{\exp\left\{-\tilde{\boldsymbol{y}}_{k,\ell}^{\Hi}\mathbf{\Sigma}_{\alpha_{\ell}}^{-1}\tilde{\boldsymbol{y}}_{k,\ell}\right\}}{(2\pi)^{\beta/2}\mathrm{det}\{\mathbf{\Sigma}_{\alpha_{\ell}}\}}
		,\end{align}
	where $\mathbf{\Sigma}_{\alpha_{\ell}}=\mathrm{diag}\{\boldsymbol{g}_{\alpha_{\ell}}\}+\sigma^2\mathbf{I}_{\beta}$.
	Thus, the pdf of $\tilde{\boldsymbol{y}}_k$ is:
	\begin{align}\label{eq.pdf_int}
		&\mathrm{Pr}(\tilde{\boldsymbol{y}}_k)=\frac{1}{2\alpha_{\mathrm{max}}+1}\sum_{\alpha_{\ell}=-\alpha_{\mathrm{max}}}^{\alpha_{\mathrm{max}}}\mathrm{Pr}(\tilde{\boldsymbol{y}}_k|\alpha_{\ell})
		,\end{align}
	given that the Doppler is uniformly distributed.
	Finally, as discussed earlier that the segments associated with different paths are mutually independent, the likelihood that $\tilde{\boldsymbol{x}}_k = \boldsymbol{x}>>k$ was transmitted can be calculated by taking the product of the pdfs of all segments averaged over their respective Doppler shifts. Then,
	%	\begin{equation}
		%		\mathcal{L}(\boldsymbol{x}_k)=\prod_{{\ell}=0}^{L-1}\mathrm{Pr}(\tilde{\boldsymbol{y}}_k).
		%	\end{equation}
	%	
	the value of $k$ with the highest likelihood will be decided as the shift applied at the transmitter:
	\begin{align}\label{eq.or_LLop}
		\bar{k} &= \underset{k=0,1\dots N-1}{\mathrm{argmax}}\; \mathcal{L}(\tilde{\boldsymbol{x}}_k)=\underset{k=0,1\dots N-1}{\mathrm{argmax}}\;\prod_{{\ell}=0}^{L-1}\mathrm{Pr}(\tilde{\boldsymbol{y}}_k). 
		%&=\underset{k=0,1\dots N-1}{\mathrm{argmin}}\quad \mathrm{max}\left\{\mathrm{IFFT}\left\{(\boldsymbol{a}_2<<k)\otimes\boldsymbol{x}\right\}\right\}.
	\end{align}
	
	\subsection{Non-integer Doppler case}
	In the non-integer Doppler case, the response of the each channel path will not be centered at an exact integer-indexed bin in the affine domain. As a result, the channel response of the $\ell$th path is spread over all samples of the $\ell$th segment, and the AFDM modulator must employ a larger spacing to account for the spreading response (nonzero $\xi$) \cite{Bemani2023}. 
	%	Similar to the integer Doppler case, with $c_1$ set in (10), $\mathbf{H}_i$ and $\mathbf{H}_j$ will not have any overlapped non-zero if $i\neq j$. 
	%	Consequently, when $\boldsymbol{x}_p$ is transmitted over the channel, the samples $y[N-Q+\alpha_{\mathrm{max}}+\nu]$ to $y[N-1]$ and $y[0]$ to $y[\alpha_{\mathrm{max}}+\nu]$ of $\boldsymbol{y}$ are associated with the pilot sample $x_p[0]=P$, which are $(\max(\tau_{\ell})+1)(2(\alpha_{\mathrm{max}}+\nu)+1)$ samples in total. Among these samples, every segment of $\gamma=2(\alpha_{\mathrm{max}}+\nu)+1$ samples are associated with a different path. 
	%	
	%	However, unlike the integer Doppler case, where each segment only contains one non-zero sample, in non-integer Doppler case, there could be from one to $2\nu$ non-zeros depending on the value of the Doppler.
	This modifies the distributions in \eqref{eq.dis_nonint} to
	\begin{align}\label{eq.19}
		%\tilde{y}_i[(-p\tilde{\alpha}+\alpha_{p})\mathrm{mod}N]&\sim \mathcal{CN}(0,\rho_p+\sigma^2)\nn\\
		\tilde{\boldsymbol{y}}_{k,\ell}&\sim \mathcal{CN}(0,\mathbf{\Psi}_{\nu_{\ell}}),\;
	\end{align}
	where $\mathbf{\Psi}_{\nu_{\ell}}=\rho_{\ell}\boldsymbol{b}_{\nu_{\ell}}\boldsymbol{b}_{\nu_{\ell}}^{\Hi}+\sigma^2\mathbf{I}_{\beta}$, and $	\boldsymbol{b}_{\nu_{\ell}}=[	{b}_{\nu_{\ell}}[0],	{b}_{\nu_{\ell}}[1]\dots	{b}_{\nu_{\ell}}[2\xi]]^{\T}$ and is given as in \cite{Bemani2023}:
	\begin{align}\label{eq.20}
		{b}_{\nu_{\ell}}[i]&=\frac{\exp\left\{-j2\pi (i-\xi-\nu_{\ell})\right\}-1}{\exp\left\{-j2\pi (i-\xi-\nu_{\ell})/N\right\}-1}\nn\\
		&\times\exp\left\{-j2\pi c_2((i-\xi)\mathrm{mod}N)^2\right\}.
	\end{align}
	\textit{Proof:} Without the loss of generality, the distribution of $\tilde{\boldsymbol{y}}_{k,\ell}$ can be calculated with $\ell=0$ for simplicity, while remaining valid for all $\ell$. As discussed earlier, $\tilde{\boldsymbol{y}}_{k,0}$ can be extracted from the first column of  $\mathbf{H}_{\mathrm{eff}}$ as shown in Fig.~\ref{fig-bin-pilot2}.Since 	$
	\mathbf{H}_{\mathrm{eff}} \triangleq \sum_{{\ell}=0}^{L-1}h_{\ell}\mathbf{H}_{\ell}
	$, and different $\mathbf{H}_{\ell}$ can be considered to have no non-zeros at the same positions \cite{Bemani2023},  $\tilde{\boldsymbol{y}}_{k,0}$ are the non-zeros elements on the first column of $h_0\mathbf{H}_0$,  defined in \cite{Bemani2023} as:
	
	\begin{align}
		H_{\ell}[p,q] &=\frac{1}{N}\exp\left\{-j{2\pi}c_2p^2\right\}\mathcal{F}_{\ell}[p,0],
	\end{align}
	where $\mathcal{F}_{\ell}[p,q]$ is defined for non-integer Doppler  in  \cite{Bemani2023} as
	\begin{align}
		\mathcal{F}_{\ell}[p,q]=
		\frac{\exp\left\{-j2\pi (p-q+2Nc_1\tau_{\ell}+\nu_{\ell})\right\}-1}{\exp\left\{-j2\pi (p-q+2Nc_1\tau_{\ell}+\nu_{\ell})/N\right\}-1}.
	\end{align}
	By substituting $\tau_0=0$, $q=0$, the non-zero elements on the first column of $\mathbf{H}_0$, which are from the $(N-\alpha_{\xi})$th to $(N-1)$th row and from $0$th to the $\alpha_{\xi}$th row can be obtained as $	\boldsymbol{b}_{\nu_{0}}$ in \eqref{eq.20}. As a result, the distribution of $\tilde{\boldsymbol{y}}_{k,0}=h_0\boldsymbol{b}_{\nu_{0}}$ can be obtained as in \eqref{eq.19}. $\hfill\blacksquare$

	Consequently,
	the pdf of  $\tilde{\boldsymbol{y}}_k$ given that the Doppler of the $\ell$-th tap is $\nu_{\ell}$ is
	\begin{align}\label{eq.pdf_nonint}
		&\mathrm{Pr}(\tilde{\boldsymbol{y}}_k|\alpha_{\ell})=\frac{\exp\left\{-\tilde{\boldsymbol{y}}_{k,\ell}^{\Hi}\mathbf{\Psi}_{\nu_{\ell}}^{-1}\tilde{\boldsymbol{y}}_{k,\ell}\right\}}{(2\pi)^{\beta/2}\mathrm{det}\{\mathbf{\Psi}_{\nu_{\ell}}\}}.
	\end{align}
	Then, the likelihood that $\tilde{\boldsymbol{x}}_k = \boldsymbol{x}>>k$ was transmitted is:
	\begin{equation}\label{eq.LLHnon_int}
		\mathcal{L}(\tilde{\boldsymbol{x}}_k)=\prod_{{\ell}=0}^{L-1}\left(\int_{-\alpha_{\mathrm{max}}-1}^{\alpha_{\mathrm{max}}+1}\mathrm{Pr}(\tilde{\boldsymbol{y}}_{k,\ell}|\nu_{\ell})\mathrm{d}\nu_{\ell}\right).
	\end{equation}
	Evaluating this likelihood function is challenging due to the integral and the complex structure of $\mathbf{\Psi}_{\nu_{\ell}}$, which is dependent on the non-integer Doppler $\nu_{\ell}$. Thus, rather than directly evaluating \eqref{eq.LLHnon_int} by solving the integral, we can instead approximate it with a Riemann sum.
	The sum can be evaluated by calculating the term  $\mathrm{Pr}(\tilde{\boldsymbol{y}}_{k,\ell}|\nu_{\ell})$	at some discrete, equally spaced frequencies of $\nu_\ell=k/G$, where $k$ takes integer values (frequency step of $1/G$, where $G$ is a positive integer). Then, the integral in \eqref{eq.LLHnon_int} can be approximated with:
	\begin{equation}\label{eq.LLH_nontin2}
		\tilde{\mathcal{L}}(\tilde{\boldsymbol{x}}_k)=\prod_{{\ell}=0}^{L-1}\left(\frac{1}{2\lambda+1}\sum_{k=-\lambda}^{\lambda}\mathrm{Pr}\left(\tilde{\boldsymbol{y}}_k\Big|\nu_{\ell}=\frac{k}{G}\right)\right),
	\end{equation}
	where $\lambda = G(\alpha_{\mathrm{max}}+1)$. By reducing the discrete frequency step (increasing $G$), the accuracy of the likelihood function can be improved, at the cost of higher complexity.
	
	Note that the MLB  detector derived above does not require CSI.  Thus, after the shift has been detected, the received signal $\boldsymbol{y}$ can be circularly shifted back to the original position, then the channel estimation can be performed normally as in \cite{Bemani2023}.
	
	\subsection{Shift candidate spacing to reduce blind estimation error}
	It can be easily seen that the performance of MLB blind estimation derived above is strongly dependent on how distinguishable the positions of the pilot are with different shifts. For example, if the two shift candidates are only one sample apart (e.g. $k=1$ and $k=2$), the resulted positions of the pilot and the guard band are only different by one sample. As a result, $\tilde{\boldsymbol{x}}_1$ and $\tilde{\boldsymbol{x}}_2$ would be very similar, 	leading to  likelihoods that are close in values. Thus, we propose to maintain a spacing of $D$ samples between different candidates to reduce error. For example, with $D=Q$, the pilot band with different candidates can only be overlapped by only a half guard ban ($Q$ samples), which makes them more recognizable. In this simulation results, the effect of this spacing will be further illustrated and discussed. Consequently, the search for the optimal shift will be modified from \eqref{eq.or_op} to:
		\begin{align}\label{eq.mod_op}
		\hat{k} =\underset{k=0,D,\dots D\times\lfloor(N-1)/D\rfloor}{\mathrm{argmin}} \mathrm{max}\left\{|\mathrm{IDFT}\left\{\boldsymbol{a}_2\otimes(\boldsymbol{x}>> k)\right\}|\right\}
		%&=\underset{k=0,1\do.
	\end{align}
	
	\section{ Complexity Analysis}\label{Sec:com}
		The proposed method searches for the shifted candidate with the lowest PAPR to transmit. Thus, with $V=\lfloor N/D\rfloor$ candidates, where $D$ is the spacing between to candidates with the unit of sample, the total complexity for the PAPR reduction at the transmitter will be $\mathcal{O}(VN\mathrm{log}_2N)$, which is $V$ times higher than the conventional AFDM with no PAPR reduction.
		
		On the other hand, the MLB detection  requires the calculation of the pdf in \eqref{eq.pdf_nonint} with the complexity of $\mathcal{O}((\alpha_{\mathrm{max}}+\xi)^2)$. Consequently, evaluating the likelihood function  in \eqref{eq.LLH_nontin2}, which requires the calculation of \eqref{eq.pdf_nonint} for $\lambda=G(\alpha_{\mathrm{max}}+1)$ discrete frequencies, $L$ paths and $V$ shift candidates requires the complexity of $\mathcal{O}(LVG(\alpha_{\mathrm{max}}+1)(\alpha_{\mathrm{max}}+\xi)^2)$.  Thus, the complexity of the receiver grows linearly with the number of taps, 	the size of the shift candidate set, and the number of discrete frequencies to calculate the likelihood functions in \eqref{eq.LLH_nontin2}; while being a quadratic function of the maximum delay and frequency spacing parameter $\xi$. However, $\alpha_{\mathrm{max}}$ and $\xi$ is usually of small values, which would not raise the complexity of the blind shift detection significantly, and hence, the complexity\textemdash no-SI trade-off can be justified.

		%	It can be seen that $\boldsymbol{\phi}_k$ is a discrete time sequence with the discrete frequency of $2kc_2N$. As a result, this sequence  will only have several dominant samples focusing around the integer frequencies closest to $2kc_2N$ and $2(k-N)c_2N$, while the remaining are of very small magnitudes. Thus, rather than performing the circular convolution between $\tilde{\boldsymbol{s}}_0$ and $\boldsymbol{\phi}_k$, which has the complexity of $\mathcal{O}(N^2)$, an approximated vector $\tilde{\boldsymbol{\phi}}_k$ can be used instead of $\boldsymbol{\phi}_k$, where  samples with magnitude below than a preset threshold $\zeta$ set to zeros to reduce the computational complexity. Suppose the number of non-zero samples in $\tilde{\boldsymbol{\phi}}_k$ is $V<<N$, the complexity of the proposed PAPR reduction search will be $\mathcal{O}(\lfloor N/D \rfloor NV)$, where $D$ is the minimum spacing between different shift candidates.
		
		\section{Simulation Results}\label{Sec:sim}
		This section provides simulation results  to illustrate the merits of the proposed   method in terms of the PAPR reduction capability and symbol error rate (SER)  with a solid-state power amplifier (SSPA) with the smoothness parameter $\gamma=3$, and the saturation (limit) amplitude $A_{\mathrm{sat}}$ varied proportionally to  the standard deviation of the AFDM time domain signal $\sigma_x$ \cite{Rapp1991}. The results are obtained  with  16 QAM,  the  Doppler $\nu_\ell\in[-1,1]$.  The parameter $c_2$ is set at $2\pi/1000$, and the spacing parameter $\xi=3$. The frequency step to calculate the likelihood function  is $1/4$ ($G=4$). The power of the pilot is set to $(2Q+1)P_{\mathrm{QAM}}$ to maintain the SNR of the time-domain signal equal to the SNR of the data on each subcarrier. The linear minimum mean squared error  equalization will be used for data detection. The PAPR reduction performance will be evaluated based on the  complementary cumulative density function (CCDF), which is defined as the probability that the PAPR is greater than  $\mathrm{PAPR}_0$, mathematicaly $\mathrm{Pr}(\mathrm{PAPR}\geq\mathrm{PAPR}_0)$. The SER and PAPR performance of the proposed method is compared with the work in \cite{Choi2026}, which recently reports the best PAPR performance, does not requires SI, and does not affect the channel estimation. According to  \cite{Choi2026},  the optimal $c_2$ values for the lowest PAPR is exhaustively searched from a set of 8 candidates $\{\pm1/N,\pm2/N,\pm3/M,\pm4/N\}$.

\begin{figure}[t!]
			\centering
			\includegraphics[width=0.5\textwidth]{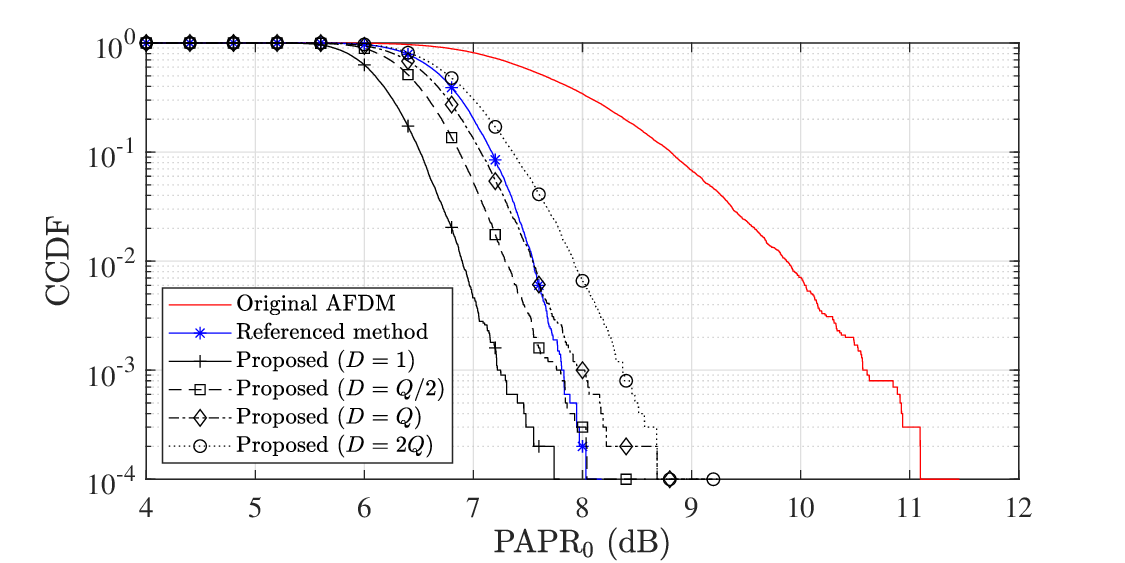}
			\caption{PAPR with different candidate spacing ($N=256$).}
			\label{fig2}
		\end{figure}

			Fig.~\ref{fig2} shows the PAPR of the proposed scheme with different candidate spacings. 	When the search is performed over all $N$ shift candidates, the proposed method can achieve up to 4 dB gain at the CCDF of $10^{-4}$, which is then reduced to 2.5 dB when the candidate spacing is expanded to $2Q$ samples. Moreover, with $D=Q/2$, the performance of the proposed method  is equivalent to that of the reference method in \cite{Choi2026}.
			
\begin{figure}[t!]
			\vspace{-0.5cm}
			\centering
			\includegraphics[width=0.5\textwidth]{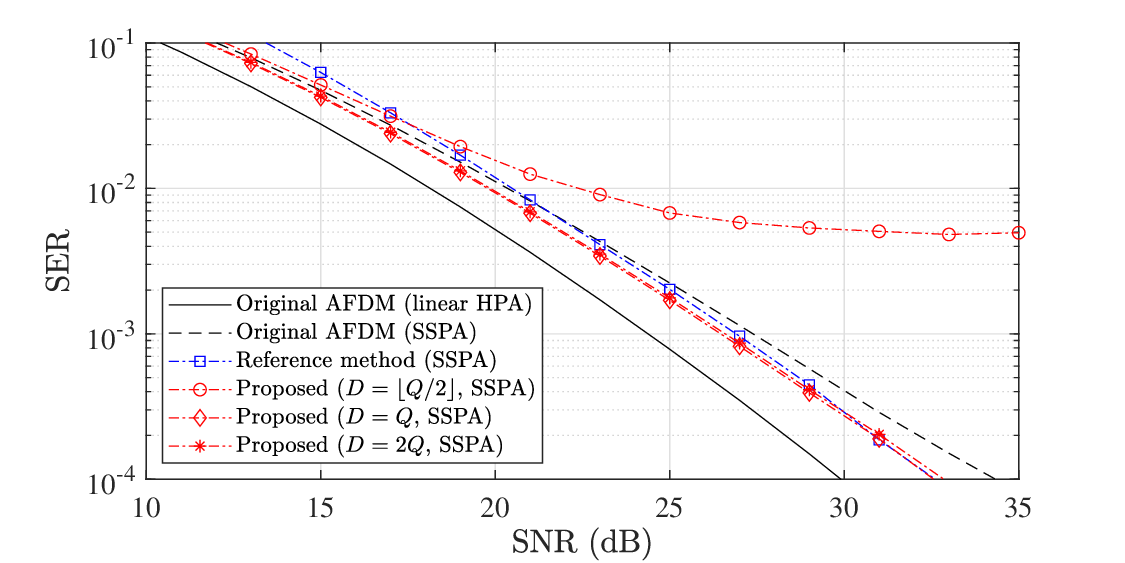}
			\caption{SER with different candidate spacing  ($N=256$, $A_{\mathrm{sat}}=1.7\sigma_{x}$).}
			\label{fig3}
		\end{figure}

\begin{figure}[t!]
			\vspace{-.5cm}
			\centering
			\includegraphics[width=0.5\textwidth]{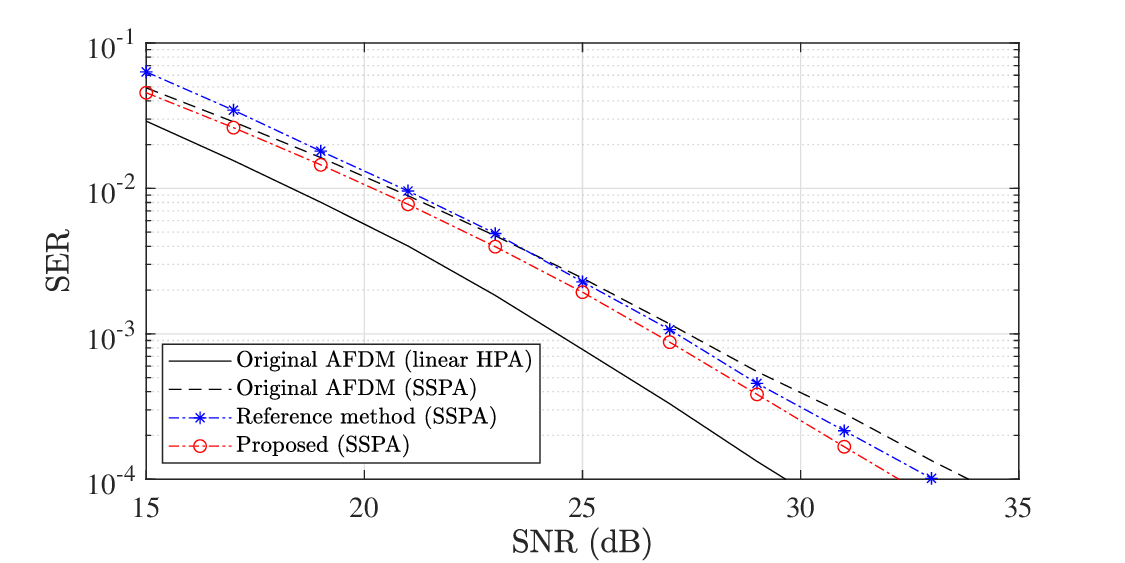}
			\caption{SER comparison with \cite{Choi2026} ($N=512$,$A_{\mathrm{sat}}=1.7\sigma_{x}$, $D=Q$).}
			\label{fig4}
		\end{figure}

			As explained earlier, when the candidates are too close, the MLB blind detection cannot distinguish how much the position of the pilot has been shifted, which leads to higher SER. This can be seen in the SER comparison with $N=256$ subcarriers in Fig.~\ref{fig3}, where a candidate spacing of $D=\lfloor Q/2\rfloor$ results in a very poor SER, reaching  an error floor due to indistinguishable candidates. On the other hand, with $D=Q$ or $2Q$, the MLB shift detector can successfully estimate the shift, which results in roughly 2 dB gain over original AFDM  without PAPR reduction 	(over SSPA) at the SER of $10^{-4}$,  equivalent to the performance of \cite{Choi2026}, while being $2.5$ dB behind original AFDM with linear HPA. 		Whereas, with $N=512$ subcarriers and $D=Q$ shown in Fig.~\ref{fig4}, the proposed method can achieve a gain of roughly 1 dB over the  method in \cite{Choi2026} at the SER of $10^{-4}$. This is due to the fact that with a higher number of subcarriers, while maintaining the same candidate spacing, the proposed method will have a bigger candidate set, which leads to improvement in PAPR, and hence, also an improvement in the SER as compared to  \cite{Choi2026}. 
		
		
		\begin{figure}[t!]
			\centering
			\includegraphics[width=0.5\textwidth]{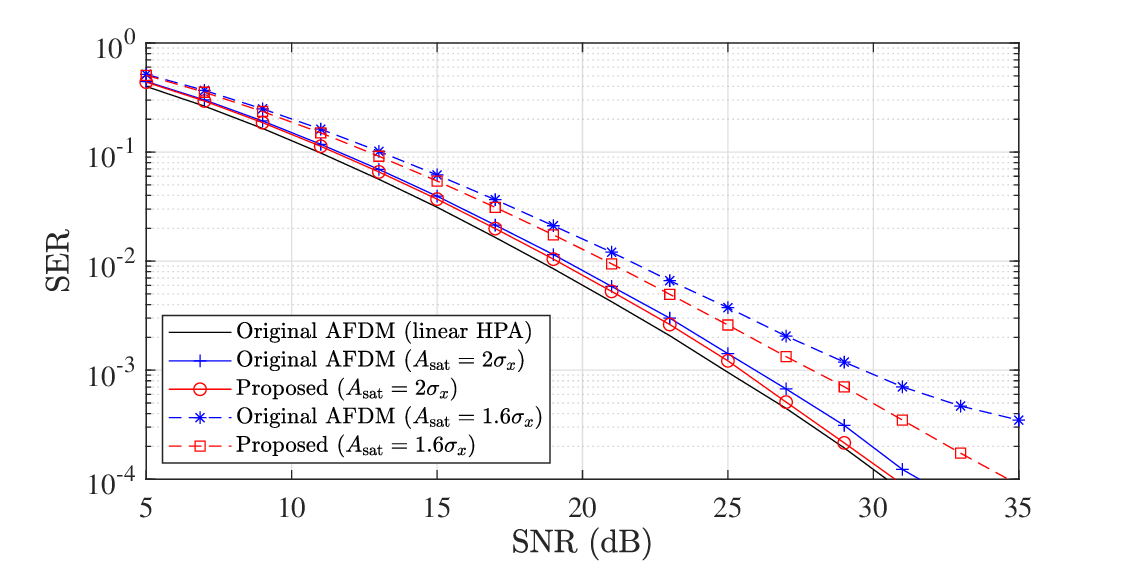}
			\caption{SER with different SSPA qualities ($N=256$, $D=Q$).}
			\label{fig5}
		\end{figure}
		Fig.~\ref{fig5} shows the SER performance when the quality of the SSPA varies. 
		When the saturation level $A_{\mathrm{sat}}=2\sigma_{x}$, the performance of the propose scheme is almost identical to that of original AFDM with linear HPA, and is roughly 1 dB better than AFDM with no PAPR reduction at the SER of $10^{-4}$. With $A_{\mathrm{sat}}=1.6\sigma_x$, this gap is significantly widen when the original AFDM is likely to approach an error floor due to the non-linear distortion introduced by the HPA. Whereas, the proposed scheme can achieve a performance which is only 3 dB worse than the  linear HPA case at the SER of $10^{-4}$. 
		%	\begin{figure}[t!]
			%		\centering
			%		\includegraphics[width=0.5\textwidth]{fig_PAPR_threshold}
			%		\caption{PAPR performance with different sample limits.}
			%		\label{fig3}
			%	\end{figure}
		%
		%\textbf{Comment on how the PAPR can be reduced to up to 3 dB as can be seen from the simulation result:} I am not sure about this yet, but I have an intuitive explanation as follows:
		%
		%Assume $2kc_2N$ is equal $5/2$. This means that $\mathrm{IFFT}\left\{\boldsymbol{\phi}_k\right\}$ will be symmetry about the frequency of $5/2$, with the value of the samples proportional to the $\mathrm{sinc}$ function. The magnitude of these sample will obviously smaller than 1. As I observed from Matlab, in some case, the peak magnitude of $\mathrm{IFFT}\left\{\boldsymbol{\phi}_k\right\}$ can reach down to 0.3. Suppose that we have a very extreme AFDM modulated signal with only one non-zero sample (PAPR is $N$ times). By convoluting with $\mathrm{IFFT}\left\{\boldsymbol{\phi}_k\right\}$, the only peak of the original AFDM signal can be reduced by  70\%. I think this might be a clue of how to choose the value of $c_2$ to improve the PAPR reduction performance. This will not work with OFDM , since $\mathrm{IFFT}\left\{\boldsymbol{\phi}_k\right\}$ will always have only one non-zero sample, which will not affect the PAPR.

		\section{Conclusions}\label{Sec:con}
		This paper proposes a PAPR reduction method for AFDM signals based on circular shift in the affine domain. The method searches for the optimal number of samples that the original affine-domain signal needs to be circularly shifted to achieve the lowest PAPR before transmission. At the receiver, a 	MLB shift detection rule is derived, which exploiting the pilot embedded structure of AFDM symbols. The proposed ML-based blind detection can estimate the shift performed at the transmitter without SI, while not affecting the pilot structure of AFDM nor requiring CSI. As a result, the channel estimation can be performed as conventional after the shift is detected. To increase the reliability of the blind shift detector, we limit the number of shift candidates by maintaining spacing between candidates to make them more distinguishable at the receiver, which also helps reduce the 	search complexity for PAPR reduction purposes. Simulation results show that the proposed scheme can achieve around 2.5 to 4 dB in PAPR reduction, which is translated to up to 3 dB gain at the SER of $10^{-4}$, depending on the SSPA quality.

		\bibliographystyle{IEEEtran}

	\end{document}